\documentclass[showpacs,twocolumn,preprintnumbers,amsmath,amssymb,pra ]{revtex4-2}

\usepackage{graphicx}
\usepackage{nicefrac}
\usepackage{color}
\usepackage{braket}
\usepackage{hyperref}

\begin{document}

\title{
Spectral and temporal properties of type-II parametric down-conversion: The impact of losses during state generation
}

\author{Denis A. Kopylov$^{1,2}$}
\email{denis.kopylov@uni-paderborn.de}
\author{Michael Stefszky$^{2,3}$}
\author{Torsten Meier$^{1,2}$ }
\author{Christine Silberhorn$^{2,3}$}
\author{Polina R. Sharapova$^{1}$}
\affiliation{
1. Department of Physics, Paderborn University, Warburger Str. 100, D-33098 Paderborn, Germany\\
2. Institute for Photonic Quantum Systems (PhoQS), Paderborn University, Warburger Str. 100, D-33098 Paderborn, Germany\\
3. Integrated Quantum Optics, Paderborn University, Warburger Str. 100, D-33098 Paderborn, Germany
}

\date{\today}

\begin{abstract}
In this paper, we theoretically study spectral and temporal properties of pulsed spontaneous parametric down-conversion (SPDC) generated in lossy waveguides.
Our theoretical approach is based on the formalism of Gaussian states and the Langevin equation, which is elaborated for weak parametric down-conversion and photon-number-unresolved click detection.
Using the example of frequency-degenerate type-II SPDC generated under pump-idler group-velocity-matching condition, we show how the joint-spectral intensity, mode structure, normalized second-order correlation function, and Hong-Ou-Mandel interference pattern depend on internal losses of the SPDC process.
In addition, we propose a new method for the experimental determination of internal losses of nonlinear waveguides which is based on the measurement of the normalized second-order correlation functions.
\end{abstract}

\maketitle

\section{Introduction}
\label{sec_intro}

Currently, for applications in quantum technologies there is a huge demand for compact integrated sources of non-classical light~\cite{Pelucchi2022}. 
One of the flexible frameworks, which allows the experimental realization of various types of non-classical field sources is based on spontaneous parametric down-conversion (SPDC).
The generation of photon pairs via SPDC requires a second-order nonlinear susceptibility; therefore, miniaturized integrated waveguide-based SPDC sources rely on the technologies for waveguide fabrication of such materials as KTP~\cite{Satyanarayan1999}, LiNbO$_3$~\cite{Sohler_2012}, or GaAs~\cite{Stanton2020}.

Nonlinear waveguides have significant benefits compared to nonlinear bulk crystals. 
The guided modes provide a high degree of localization of the electromagnetic field~\cite{Agrawal_book}, effective coupling between the pump, signal, and idler fields, and the tunability of their dispersion by the geometry of the waveguide~\cite{Mishra22}.
However, imperfections during waveguide fabrication result in differences between the desired ideal and the fabricated waveguide~\cite{Melati_2014} which may lead to a change of the properties of the generated states. 
Importantly for quantum technological applications is the determination and characterization of internal waveguide losses during SPDC. For example, signal and idler photons can be scattered due to the roughness of the waveguide surface~\cite{Hammer_2024}. 
In turn, AlGaAs waveguides, which are also used for SPDC~\cite{Placke_2024,Schuhmann_2024_PRX}, have a strong material absorption in their cores.
Therefore, the proper description of such non-ideal lossy SPDC sources is a relevant task.
 
SPDC sources may be characterized using one or more of several experimental techniques.
Measurements of the joint spectral intensity, the normalized second-order correlation function, and the Hong-Ou-Mandel interference~\cite{Hong_1987} represent standard tools~\cite{Avenhaus09,Christ_2011,Zielnicki_2018,Graffitti2018,Xin_2022,Lange_2023}.
They have been shown to be convenient for bulk crystals and are also widely used for the experimental characterization of lossy waveguide sources.
However, the standard description and interpretation of experimental results does not take into account the presence of internal losses.

In this paper, we highlight the fundamental difference between the pulsed SPDC generated in media with and without internal losses and present a method to experimentally indicate the presence of internal losses and measure their amount.

The structure of this paper is the following: In Section~\ref{sec_theory_equations}, we present our theoretical approach which is based on the framework of Gaussian states and the Langevin equation. The generated PDC state is described in terms of the second-order correlation matrices, and in Section~\ref{sec_theory_spectral_temporal}, we present explicit expressions for the joint spectral intensity and the temporal profiles of the signal and idler fields.
In Section~\ref{sec_theory_modes}, we present the Mercer-Wolf-basis and the number of occupied modes for type-II SPDC.
Sections~\ref{sec_theory_hom} and~\ref{sec_theory_g2} show how the HOM interference pattern and the normalized second-order correlation functions can be computed for Gaussian states. 
In Section~\ref{sec_theory_sum}, we summarize the advantages of our method.   
Section~\ref{sec_results} presents and discussed the results of numerical simulations of frequency-degenerate type-II SPDC generated under the pump-idler group-velocity-matching condition. 
The obtained results allow us to propose a new method for experimental determination of internal loss coefficients from the measured normalized second-order correlation functions, as is presented in Sec~\ref{sec_results_new_method}.

\section{Theoretical approach}
\label{sec_theory}

\subsection{Master equation for type-II parametric down-conversion}
\label{sec_theory_equations}

\begin{figure}
    \includegraphics[width=1.\linewidth]{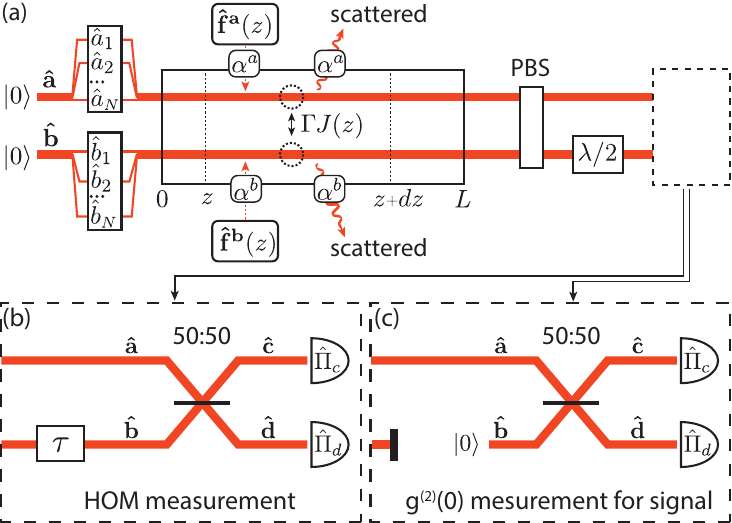 } 
    \caption{ 
        (a) The PDC generation scheme in lossy media; 
        (b) the Hong-Ou-Mandel interference scheme; 
        (c) scheme for measuring the normalized second-order correlation function for the signal field $g_s^{(2)}$.
        }
    \label{fig_1_scheme}
\end{figure}

For the numerical analysis of parametric down-conversion (PDC) with internal losses, we use the numerical scheme that was developed in~\cite{Kopylov_2024}.
The approach is based on the framework of multimode Gaussian states~\cite{Weedbrook_2012} in a discrete uniform frequency space $(\omega_0, \omega_1, \dots, \omega_N)$, which allows us to write the equations of motion directly in the form which are used in our numerical calculations.
For type-II PDC, the nonlinear interaction produces two orthogonally polarized fields: horizontal (here TE) and vertical (TM) polarized field components. 
Further in the text, we call these field components signal and idler, respectively.
Signal and idler fields at position $z$ are given by two vectors of monochromatic operators:
$\hat{\mathbf{a}}(z) = \big(\hat{a}_0(z), \hat{a}_1(z), \dots, \hat{a}_N(z)\big)^T$ and $\hat{\mathbf{b}}(z) = \big(\hat{b}_0(z), \hat{b}_1(z), \dots, \hat{b}_N(z)\big)^T$, respectively, where $\hat{a}_n(z) \equiv \hat{a}(z,\omega_n)$ and $\hat{b}_n(z) \equiv \hat{b}(z,\omega_n)$.
These operators obey bosonic commutation relations $[\hat{a}_i(z), \hat{a}^\dagger_j(z)]=[\hat{b}_i(z), \hat{b}^\dagger_j(z)]=\delta_{ij} $ and $[\hat{a}_i(z), \hat{b}^\dagger_j(z)]= 0$.
In terms of fast oscillating operators, the electric field operator for the signal field has the form
\begin{equation}
\hat{E}_a^+(z, t) = \sum_{m} \xi_a(\omega_m)  \hat{a}_m(z) e^{- i \omega_m t},
\end{equation} 
where the amplitude $\xi_a(\omega_m) = \sqrt{\frac{\hbar \omega_m}{2 \varepsilon_0 c T n^a(\omega_m)} }$, $T = \frac{2\pi}{\omega_{m+1}-\omega_{m}}$, and $n^a(\omega_m)$ is the refractive index for the signal field. For the idler field the index `$a$' should be replaced by `$b$'. 

The generator of the spatial evolution~\cite{Huttner_1990,Horoshko_2022} for type-II PDC is given by $\hat{G}(z)=\hat{G}_{l}(z)+\hat{G}_{pdc}(z)$, where the linear part is given by
\begin{multline}
\hat{G}_{l}(z) =  \sum_{n}  \hbar k^a_n \hat{a}_n^\dagger(z) \hat{a}_n(z) \\ + \sum_{n}  \hbar k^b_n \hat{b}_n^\dagger(z) \hat{b}_n(z) + h.c. 
\label{eq_ap_momentum_lin}
\end{multline} 
and the nonlinear interaction part is 
\begin{equation}
\hat{G}_{pdc}(z) =   \dfrac{\hbar\Gamma}{2} \sum_{i,j} J_{ij}(z)  \hat{a}_i^\dagger(z) \hat{b}_j^\dagger(z)  + h.c.
\label{eq_ap_momentum_nonlin}
\end{equation} 
Here the coupling matrix $J_{ij}(z) = S(\omega_i + \omega_j) e^{i (k_p(\omega_i + \omega_j) - k_{QPM})z } $; $S(\omega)$ is the pump spectrum at $z=0$;
$k^{(a,b,p)}_n \equiv k^{(a,b,p)}(\omega_n) = \frac{n^{(a,b,p)}(\omega_n)\omega_n}{c}$ are the wavevectors of the (a) signal, (b) idler, and (p) pump fields in a waveguide; $k_{QPM} = 2\pi/\Lambda$ and $\Lambda$ is the poling period for the quasi-phase-matching condition. 
The parameter $\Gamma$ determines the coupling strength and is in the spontaneous regime of PDC for $\Gamma \ll 1 $.
However, note that  all further equations are also valid  for arbitrarily large $\Gamma$. 

As we are interested in internal PDC losses, i.e., losses during the PDC generation, we need to describe the dynamics in terms of an open quantum system~\cite{Vogel_Welsch_book}. 
For simplicity, we introduce two separate, non-interacting, spatially delta-correlated Markovian environments for the signal and idler modes, which allow us to introduce two sets of Langevin noise operators $\hat{f}^a_n(z) \equiv \hat{f}^a(z, \omega_n)$ and $\hat{f}^b_n(z) \equiv \hat{f}^b(z, \omega_n)$ and two frequency-independent loss-coefficients $\alpha_a$ and $\alpha_b$ (see Fig.~\ref{fig_1_scheme}(a)). 
The spatial Langevin equation for the operators $\hat{a}$ has the form~\cite{Kopylov_2024}
\begin{multline}
    \dfrac{d \hat{a}_n(z) }{d z} = i \kappa^{a}_n \hat{a}_n(z)  \\ + i\Gamma \sum_{m} J_{nm}(z) \hat{b}^\dagger_m(z)  + \sqrt{\alpha_a } \hat{f}^a_n(z),   
    \label{eq_langevin}
\end{multline}
where $\kappa^a_n = k^a_n+i\alpha_a/2$. 
The Langevin equation for operators $\hat{b}$ is similar.
In the absence of losses, the Langevin equation corresponds to the spatial Heisenberg equation~\cite{Horoshko_2022}.

In contrast to the lossless case, where the solution to the Heisenberg equation has the form of a Bogoliubov transformation~\cite{Christ_2013,
Sharapova_2020}, the solution to the multimode Langevin equation (Eq.~\eqref{eq_langevin}) does not have such a simple form.  
However, in this paper, we consider the case where the initial state and environment are given by vacuum states, which leads to the generation of an undisplaced Gaussian states via the PDC process.
Therefore, the spatial evolution of PDC light is described by a master equation for the second-order correlation functions~\cite{Kopylov_2024}.
In a discrete frequency space, the master equation constitutes a system of differential equations. 
To write this system in a compact matrix form, we introduce the second-order correlation matrices $\mathcal{D}(z)$ and $\mathcal{C}(z)$ as
\begin{equation}
    \mathcal{D}(z) =     
    \begin{pmatrix}
        \braket{\mathbf{\hat{a}^\dagger \hat{a}} }_z   & \braket{\mathbf{\hat{a}^\dagger \hat{b}} }_z   \\
        \braket{\mathbf{\hat{b}^\dagger \hat{a}} }_z   & \braket{\mathbf{\hat{b}^\dagger \hat{b}} }_z  
    \end{pmatrix}, ~ 
    \mathcal{C}(z) =     
    \begin{pmatrix}
        \braket{\mathbf{\hat{a} \hat{a}} }_z   & \braket{\mathbf{\hat{a} \hat{b}} }_z   \\
        \braket{\mathbf{\hat{b} \hat{a}} }_z   & \braket{\mathbf{\hat{b} \hat{b}} }_z  
    \end{pmatrix}.
    \label{eq_correlation_matrices_DC}
\end{equation}
The expressions in the form $\braket{\mathbf{\hat{a}^\dagger \hat{b}} }_z$ and $\braket{\mathbf{\hat{a} \hat{b}} }_z$ denote the $N \times N$ matrices with matrix elements $\braket{\hat{a }_i^\dagger(z) \hat{b}_j(z)}$ and $\braket{\hat{a}_i (z) \hat{b}_j(z)}$, respectively. 
The resulting master equation in a matrix form reads~\cite{Kopylov_2024}   
\begin{multline}
    \dfrac{d \mathcal{D}(z)}{d z} =  i \big(\mathcal{D}(z) K - K \mathcal{D}(z)\big)  \\ + i \Gamma \big(\mathcal{C}^{*}(z) M^T(z) -  M^*(z) \mathcal{C}(z) \big),
    \label{eq_master_equation_1}
\end{multline}
\begin{multline}
    \dfrac{d \mathcal{C}(z) }{d z} = i \big(\mathcal{C}(z) K + K \mathcal{C}(z)\big) \\ + i\Gamma   \Big( \big(M(z) \mathcal{D}(z)+M(z) \big)^T + M(z) \mathcal{D}(z)  \Big),
    \label{eq_master_equation_2}
\end{multline}
where the superscript $[\cdot]^*$ denotes the complex conjugation of a matrix. 
The matrix $K$ is a diagonal matrix with elements $\textrm{diag}(\kappa^a_0, \dots \kappa^a_N, \kappa^b_0, \dots, \kappa^b_N)$, while 
the $z$-dependent coupling matrix $M(z)$ is given by
\begin{equation}
    M(z)=
    \begin{pmatrix}
        \mathbf{0}_{N}       & J(z)  \\
        J^{T}(z)  &   \mathbf{0}_{N} 
    \end{pmatrix}.
    \label{eq_coupling_matrix}
\end{equation}
The initial condition (vacuum  state) reads $\mathcal{D}(0)=\mathcal{C}(0)=\mathbf{0}_{2N}$ and,
together with the coupling matrix in the form of Eq.~\eqref{eq_coupling_matrix}, determines the structure of the solution: i.e., for any $z$  the following equalities are fulfilled
$
    \braket{\mathbf{\hat{a}^\dagger \hat{b}} }_z = 
    \braket{\mathbf{\hat{b}^\dagger \hat{a}} }_z = 
    \braket{\mathbf{\hat{a} \hat{a}} }_z = 
    \braket{\mathbf{\hat{b} \hat{b}} }_z =\mathbf{0}_{N} 
$.
Therefore, the correlation matrices for type-II PDC have the form
\begin{equation}
    \mathcal{D}(z) =     
    \begin{pmatrix}
        \braket{\mathbf{\hat{a}^\dagger \hat{a}} }_z  & \mathbf{0}_{N}  \\
        \mathbf{0}_{N}  & \braket{\mathbf{\hat{b}^\dagger \hat{b}} }_z 
    \end{pmatrix}, ~ 
    \mathcal{C}(z) =     
    \begin{pmatrix}
        \mathbf{0}_{N}  & \braket{\mathbf{\hat{a} \hat{b}} }_z  \\
        \braket{\mathbf{\hat{b} \hat{a}} }_z  & \mathbf{0}_{N} 
    \end{pmatrix}.
    \label{eq_solution_block}
\end{equation}

By solving the master equations (Eqs.~\eqref{eq_master_equation_1},\eqref{eq_master_equation_2}) from  $z=0$ till $z=L$, where $L$ is the length of the nonlinear waveguide, the output second-order correlation matrices $\mathcal{D}(L)$ and $\mathcal{C}(L)$ are evaluated. 
These matrices contain all information about the quantum state. 
In the next sections, we show how these matrices can be used to compute spectral and temporal profiles of the signal and idler fields, the joint spectral intensity, and the effective number of occupied modes. 
In Sections~\ref{sec_theory_hom} and~\ref{sec_theory_g2} the correlation matrices are used to calculate the Hong-Ou-Mandel interference and the normalized second-order correlation functions.

\subsection{Spectral and temporal properties of PDC}
\label{sec_theory_spectral_temporal}

The spectral photon-number distribution for the signal field is obtained from the diagonal elements of the matrix $\braket{\mathbf{\hat{a}^\dagger \hat{a}} }_L$ as
\begin{equation}
    \braket{\hat{n}_a(\omega_m)} \equiv \braket{\hat{a}_m^\dagger(L) \hat{a}_m(L)}
\end{equation}
and defines the total number of photons in the signal subsystem $N_a = \sum_m \braket{\hat{n}_a(\omega_m)}$.
In addition to the spectral distribution, the temporal profile of the signal field at $z=L$ can be found as
\begin{multline}
    I_a(t) = \braket{\hat{E}_a^-(L, t)\hat{E}_a^+(L, t)}  \\ = \sum_{nm} \xi_a (\omega_n) \xi_b(\omega_m) \braket{ \hat{a}_n^\dagger(L) \hat{a}_m(L)}e^{ i (\omega_n-\omega_m) t}.
\end{multline}
For the idler field, the spectral and temporal intensity profile can be found in a similar by changing $a$ to $b$.

It is a little more difficult to express the fourth-order moments in terms of the second-order matrices. 
In particular, one can define the joint spectral intensity (JSI) as 
\begin{equation}
    \mathrm{JSI}(\omega_n,\omega_m) =  \braket{\hat{n}_a(\omega_n)\hat{n}_b(\omega_m)}.
    \label{eq_JSI}
\end{equation}
In order to express the $\mathrm{JSI}(\omega_n,\omega_m)$ in terms of second-order correlations, the result derived in Ref.~\cite{Phillips_2019} is used and reads
\begin{multline}
    \mathrm{JSI}(\omega_n,\omega_m) = 
    \braket{\hat{a}^\dagger_n\hat{b}^\dagger_m}\braket{\hat{a}_n\hat{b}_m} \\  +
    \braket{\hat{a}^\dagger_n\hat{a}_n}\braket{\hat{b}^\dagger_m\hat{b}_m} +
    \braket{\hat{a}^\dagger_n\hat{b}_m}\braket{\hat{b}^\dagger_m\hat{a}_n}. 
    \label{eq_JSI_detailed} 
\end{multline}

\subsection{Mode structure of PDC}
\label{sec_theory_modes}

In order to study the mode structure of the resulting fields, we use the broadband Mercer-Wolf modes~\cite{Wolf_1982,Mandel_Wolf_book}.
These modes are nothing more than a diagonalization of the matrix $\mathcal{D}$ with the use of a unitary matrix $V$~\cite{Kopylov_2024}.
As long as the matrix $\mathcal{D}$ has a block-diagonal form (see Eq.~\eqref{eq_solution_block}), $V = V_a \oplus V_b$ holds, where $V_a$ and $V_b$ are also unitary matrices.  
Therefore, for type-II PDC, the Mercer-Wolf expansion diagonalizes the signal and idler subsystems independently, allowing us to introduce broadband modes for the signal and idler subsystems separately, namely,
$\mathbf{\hat{A}} = V_a^T\hat{\mathbf{a}}$ and $\mathbf{\hat{B}} = V_b^T\hat{\mathbf{b}}$, respectively.
As a result, the correlation matrix $\mathcal{D}^{MW}$ in the broadband Mercer-Wolf mode basis has the form 
\begin{equation}
   \mathcal{D}^{MW}  = V^\dagger  \mathcal{D}  V = 
    \begin{pmatrix}
        \braket{\mathbf{\hat{A}^\dagger \hat{A}} }   & \mathbf{0}_{N}  \\
        \mathbf{0}_{N}      & \braket{\mathbf{\hat{B}^\dagger \hat{B}} } 
    \end{pmatrix},
   \label{eq_Mercer}
  \end{equation}
where both the matrices $\braket{\mathbf{\hat{A}^\dagger \hat{A}} }$ and $\braket{\mathbf{\hat{B}^\dagger \hat{B}} }$ are diagonal.

For an arbitrary correlation matrix $\mathcal{D}$, the number of occupied modes is defined as~\cite{Fabre_2020}
\begin{equation}
    \mu(\mathcal{D}) = \Big( \sum_i \big[n_i(\mathcal{D})\big] ^2 \Big)^{-1},
    \label{eq_number_of_modes}
\end{equation}
where $n_i(D)=\mathcal{D}_{ii}/(\sum_i \mathcal{D}_{ii})$.

The total effective number of Mercer-Wolf PDC modes (signal and idler) is given by
\begin{equation}
    \mu_{ab} \equiv \mu(\mathcal{D}^{MW}).
        \label{eq_number_of_modes_MW}
\end{equation}
Note that the matrix $\mathcal{D}^{MW}$ being diagonal implies that the above expression gives the minimal number of occupied modes compared to any other broadband basis~\cite{Kopylov_2024}. 
The number $\mu_{ab}$ is the total effective number of modes and, therefore, is different to the Schmidt number (the effective number of spectral modes), which is commonly defined via the Schmidt decomposition of the two-photon amplitude~\cite{Eckstein_2011, Dyakonov_2015}.

In addition, due to the fact that the Mercer-Wolf expansion diagonalizes the signal and idler subsystems independently,
an effective number of occupied Mercer-Wolf modes can be defined separately for the signal and idler subsystems as
\begin{equation}
 \mu_a \equiv  \mu(\braket{\mathbf{\hat{A}^\dagger \hat{A}} })  ~ ~\text{and} ~ ~ \mu_b \equiv \mu(\braket{\mathbf{\hat{B}^\dagger \hat{B}} }) , 
            \label{eq_number_of_modes_a_b}
\end{equation}
respectively.

Note that the Mercer-Wolf expansion for lossless {type-II} PDC leads to a diagonal correlation matrix $\braket{\mathbf{\hat{A} \hat{B}} }$, which indicates the pairwise correlations between signal and idler modes; the number of modes for the signal and idler subsystems are equal $\mu_{a}=\mu_{b}$ and $\mu_{ab}=\mu_{a}+\mu_{b}$.
In the presence of losses, non-diagonal terms appear in $\braket{\mathbf{\hat{A} \hat{B}} }$ which indicate the presence of field correlations between Mercer-Wolf modes with different indexes.
In turn, the number of modes in the signal and idler subsystems can differ ($\mu_{a}\neq\mu_{b}$).
In this case, the function $\mu(\cdot)$ is not additive, i.e., $\mu_{ab}\neq \mu_{a}+\mu_{b}$. 
Therefore, to fully characterize a lossy PDC system, all three numbers of modes, $\mu_{ab}$, $\mu_a$, and $\mu_b$, are required.






\subsection{The Hong-Ou-Mandel interferometer}
\label{sec_theory_hom}

Usually, a two-photon approximation of SPDC is used for numerical simulations of the HOM effect~\cite{Grice_1997,Branczyk_2017_arxiv}. 
The SPDC state is assumed to be pure and is characterized by the two-photon amplitude (TPA), which determines the coincidence probability in the HOM experiment.
In this paper, we consider mixed states: Due to losses, the PDC photons are generated both from initial vacuum fluctuations and from an uncorrelated environment.
Note that in the framework of Gaussian states, all information about the PDC state is contained in the correlation matrices~$\mathcal{D}$ and $\mathcal{C}$, so
 all Fock-state contributions are included in the correlation matrices, which takes our approach beyond the two-photon approximation. 

A numerical procedure to obtain the HOM interference can be split into two parts: a linear transformation of the PDC field and a detection via photon-click (on-off) detectors.
 
\subsubsection{Linear transformation for HOM}

The scheme of the HOM interferometer is shown in Fig.~\ref{fig_1_scheme}(a,b).
At the output of the waveguide, the signal and idler fields have orthogonal polarizations (Fig.~\ref{fig_1_scheme}(a)). 
Thus, a polarizing beam-splitter is used for the spatial separation of the signal and idler beams. 
To let the fields interfere at a beamsplitter, a half-wave-plate in the idler channel is used to match the polarizations of the signal and idler fields.
Note that these two optical elements keep the matrices $\mathcal{D}$ and $\mathcal{C}$ unchanged.

Varying the distinguishability in HOM interference is usually achieved by adjusting the time delay between the signal and idler fields interfering on a 50:50 beamsplitter (Fig.~\ref{fig_1_scheme}(b)).
Both these elements are described by unitary transformations of annihilation operators. 
The time delay $\tau$ is introduced for the idler field via the diagonal unitary transformation $\hat{b}_n \rightarrow \hat{b}_n e^{i\omega_n \tau}$.
The transformation for 50:50 beamsplitter is given by 
    $\hat{c}_n = \frac1{\sqrt{2}} \big(\hat{a}_n + \hat{b}_n \big)$,  $\hat{d}_n = \frac1{\sqrt{2}} \big(\hat{a}_n - \hat{b}_n \big)$,
where $\hat{c}_n$ and $\hat{d}_n$ are the output annihilation operators (see Fig.~\ref{fig_1_scheme}(b)).
In matrix form, such input-output relation reads 
\begin{equation}
    \begin{pmatrix}
        \mathbf{\hat{c}}  \\
        \mathbf{\hat{d}} 
    \end{pmatrix} = 
    \mathcal{U}(\tau)    
    \begin{pmatrix}
        \mathbf{\hat{a}}  \\
        \mathbf{\hat{b}} 
    \end{pmatrix},
\end{equation}
where 
\begin{equation}
    \mathcal{U}(\tau) =     \frac1{\sqrt{2}}
    \begin{pmatrix}
        \mathbf{1}_N  &  \mathbf{1}_N  \\
        \mathbf{1}_N  & -\mathbf{1}_N
    \end{pmatrix}
    \begin{pmatrix}
        \mathbf{1}_N  & 0  \\
        0 & V(\tau) 
    \end{pmatrix}, 
    \label{eq_unitary_transform}
\end{equation}
$\mathbf{1}_N $ is the identity matrix and  $V(\tau) = \textrm{diag}( e^{i\omega_1 \tau}, \dots ,  e^{i\omega_N \tau})$ is the diagonal matrix.

Having the unitary transformation $\mathcal{U}(\tau)$ for operators, the second-order correlation matrices $\mathcal{D}$ and $\mathcal{C}$ are transformed as~\cite{Kopylov_2024}
 \begin{equation}
    \mathcal{F} = \mathcal{U}^*(\tau) \  \mathcal{D} \ \mathcal{U}^T(\tau), ~ ~ \mathcal{E} = \mathcal{U}(\tau) \ \mathcal{C} \ \mathcal{U}^T(\tau),
    \label{eq_correlation_transformation}
\end{equation}
where resulting correlation matrices are
\begin{equation}
    \mathcal{F}(z) =     
    \begin{pmatrix}
        \braket{\mathbf{\hat{c}^\dagger \hat{c}} }_z  & \braket{\mathbf{\hat{c}^\dagger \hat{d}} }_z  \\
        \braket{\mathbf{\hat{d}^\dagger \hat{c}} }_z  & \braket{\mathbf{\hat{d}^\dagger \hat{d}} }_z
    \end{pmatrix}, ~ 
    \mathcal{E}(z) =     
    \begin{pmatrix}
        \braket{\mathbf{\hat{c} \hat{c}} }_z  & \braket{\mathbf{\hat{c} \hat{d}} }_z  \\
        \braket{\mathbf{\hat{d} \hat{c}} }_z  & \braket{\mathbf{\hat{d} \hat{d}} }_z    
    \end{pmatrix}.
\end{equation}

\subsubsection{Photon-click detectors}
For the HOM interferometer, we use two frequency-non-resolving photon-click detectors (on-off detectors), placed in both the signal and idler channels (Fig.~\ref{fig_1_scheme}(b)).
This type of detector does not distinguish the number of detected photons and their frequencies and is commonly used in HOM experiments~\cite{Babel23}.

Consider a state $\hat{\rho}$ which consists of two subsystems $c$ and $d$.
The detection operator for the subsystem $i$ reads 
\begin{equation}
    \hat{\Pi}_i = \hat{I} - \ket{\mathbf{0}}\bra{\mathbf{0}}_i,
    \label{eq_povm}
\end{equation}
where $i \in [c,d]$ and $\ket{\mathbf{0}}_i   = \bigotimes_{n} \ket{0}_i $ is a vacuum state for the $i$-th subsystem and $\hat{I}$ is the identity operator.
Then the probability of click detection in channel $c$ is 
\begin{equation}
    P_{c} =  \mathrm{Tr}\big(\hat{\Pi}_c \otimes \hat{I}_d \ \hat{\rho}\big) = 1 - q_c,
    \label{eq_prob_single}
\end{equation}
where 
\begin{equation}
     q_c = \mathrm{Tr}\big(\ket{\mathbf{0}}\bra{\mathbf{0}}_c \otimes \hat{I}_d \ \hat{\rho}\big) = \braket{\mathbf{0} |\hat{\rho}_c|\mathbf{0}}_c
     \label{eq_vac_single}
\end{equation}
and the matrix $\hat{\rho}_c = \mathrm{Tr}_d(\hat{\rho})$  is the density matrix for subsystem $c$.
The expression for the click-detection probability in channel $d$ is similar to channel $c$. 

The coincidence probability of photon-click detection in both channels reads
\begin{equation}
    P_{cd} = \textrm{Tr} \big(\hat{\Pi}_c \otimes \hat{\Pi}_d \hat{\rho}\big) = 1+q_{cd}-q_{c}-q_{d},
    \label{eq_prob_coinc}
\end{equation}
where 
\begin{equation}
    q_{cd} = \textrm{Tr} \Big( \ket{\mathbf{0}}\bra{\mathbf{0}}_c \otimes \ket{\mathbf{0}}\bra{\mathbf{0}}_d   \hat{\rho} \Big) = \bra{\mathbf{0}} \hat{\rho} \ket{\mathbf{0}}  
    \label{eq_vac_coinc}
\end{equation}
 is a probability of simultaneous detection of  vacuum  in both channels; $\ket{\mathbf{0}} = \ket{\mathbf{0}}_c \otimes \ket{\mathbf{0}}_d$.  

The Eqs.~\eqref{eq_vac_single} (and~\eqref{eq_vac_coinc}) are nothing more than the fidelity between the states $\hat{\rho}_c$ (and $\hat{\rho}$) with the vacuum states $\ket{\mathbf{0}}_c$ (and $\ket{\mathbf{0}}$).
For multimode Gaussian states these fidelities  can be expressed in terms of the covariance matrix ($\hbar=2$)~\cite{Spedalieri_2012,Banchi_2015}
\begin{equation}
    q_c = F(\sigma_c), ~ ~  q_{cd} = F(\sigma_{cd}).
\end{equation}
where 
\begin{equation}
    F(\sigma) = \dfrac{2^{M}}{\sqrt{\mathrm{det}(\sigma + \mathbf{1}_{2M})}},
\end{equation} 
$2M$ is the dimension of the covariance matrix $\sigma$, and $\mathbf{1}_{2M}$ is a $2M\times 2M$ identity matrix.
In Appendix~\ref{appendix_covariance_matrix} the equations for the covariance matrix are given explicitly.
Similar results for probabilities can be obtained via the Torontian function, which was used in Gaussian boson-sampling with threshold detectors~\cite{Quesada_2018}.

\subsection{Normalized second-order correlation function}
\label{sec_theory_g2}

The normalized second-order correlation function $g^{(2)}$ reveals additional temporal properties of the generated state.
By definition~\cite{Glauber_1963}, the normalized second-order correlation function reads
\begin{equation}
    g^{(2)}(t_1, t_2, t_3, t_4) = \dfrac{ \braket{ ~ \displaystyle \prod_{i=1}^2 \hat{E}^{(-)}(t_i) \displaystyle \prod_{j=3}^4  \hat{E}^{(+)}(t_j) ~ } }{ \displaystyle \prod_{i=1}^4 \sqrt{G^{(1)}(t_i)} },
    \label{eq_g2func_def}
\end{equation}
where $G^{(1)}(t_i) = \braket{\hat{E}^{(-)}(t_i)\hat{E}^{(+)}(t_i)}$ is the first-order correlation function.

For pulsed multimode optical fields,
the measurement of the normalized second-order correlation function in the form of Eq.~\eqref{eq_g2func_def} is quite challenging. 
Indeed, for short pulses, electric field fluctuations inside the pulse are present, which requires the use of 
nonlinear optical effects for a complete $g^{(2)}$ measurement, e.g., two-photon absorption or second harmonic generation~\cite{Boitier_2013,Kopylov_2020}, which is problematic for weak optical fields.
Usually, such fields are measured via the photon counting detectors, whose detection times are much larger than the pulse duration.
Such detectors cannot resolve the fast field fluctuations, however, the averaged $g^{(2)}$ value is usually used for an estimation of the number of PDC spectral modes ~\cite{Christ_2011} can be obtained.

In this paper, we consider the measurement of $g^{(2)}$ via the coincidence probability in a coincidence scheme with frequency-unresolved click detectors. 
The scheme is depicted in Fig.~\ref{fig_1_scheme}(c) and the normalized second-order correlation function for the signal field is given by
\begin{equation}
     g_s^{(2)} = \dfrac{P^s_{cd}}{P^s_{c}P^s_{d}},
     \label{eq_g2func_meas}
\end{equation}
where $P^s_{cd}$ is the coincidence probability and $P^s_{c}$ and $P^s_{d}$ are the detection probabilities in the signal and idler channels, respectively.
Further in the text, we use the definition Eq.~\eqref{eq_g2func_meas} to determine the normalized second-order correlation function.

To compute $g_s^{(2)}$ for the signal field according to  Eq.~\eqref{eq_g2func_meas},  we block the idler field and insert a new vacuum field instead.
Then, the state before the beamsplitter is given by the correlation matrices
\begin{equation}
    \mathcal{D}(z) =     
    \begin{pmatrix}
        \braket{\mathbf{\hat{a}^\dagger \hat{a}} }_z  & \mathbf{0}_N  \\
        \mathbf{0}_N  & \mathbf{0}_N 
    \end{pmatrix}, ~ 
    \mathcal{C}(z) =     
    \begin{pmatrix}
        \braket{\mathbf{\hat{a} \hat{a}} }_z  &  \mathbf{0}_N  \\
        \mathbf{0}_N         &  \mathbf{0}_N 
    \end{pmatrix}.
\end{equation}
The state after the beamsplitter is given by the transformation Eq.~\eqref{eq_correlation_transformation} with $\mathcal{U}(\tau=0)$.   
From Eq.~\eqref{eq_prob_single} the probabilities $P^s_{c}$ and $P^s_{d}$ are calculated and from Eq.~\eqref{eq_prob_coinc} -- the coincidence probability $P^s_{cd}$.

According to Refs.~\cite{Christ_2011,Sh_Iskhakov_2012}, the $g^{(2)}_s$ value and the number of modes $\mu_a$ for the signal field are related as
\begin{equation}
    g^{(2)}_s = 1 + 1/\mu_a.
    \label{eq_g2func_modes}
\end{equation}
For $\mu_a \geq 1$,  $g_s^{(2)} \leq 2$, while the equality holds for the spectrally single-mode regime. 
The $g_i^{(2)}$ for the idler field is calculated in a similar manner as for the signal field.

\subsection{Summary}
\label{sec_theory_sum}

Before presenting the numerical results, we emphasize the advantages of our description compared to standard approaches.

As was mentioned in Section~\ref{sec_theory_hom}, lossless low-gain PDC is usually described by first-order perturbation theory with the use of the two-photon amplitude (TPA).
In the case of pulsed low-gain PDC with losses, the correct description in terms of the TPA is quite difficult. 
The existing approaches are developed either for PDC with a monochromatic pump~\cite{Antonosyan_2014,Vavulin_2017,Gr_fe_2017} or with the use of scattering theory~\cite{Helt_2015,Poddubny_2016,Banic_2022}, whose application is challenging for long single-path waveguides and pulsed light.
\begin{figure*} 
    \includegraphics[width=0.245\linewidth]{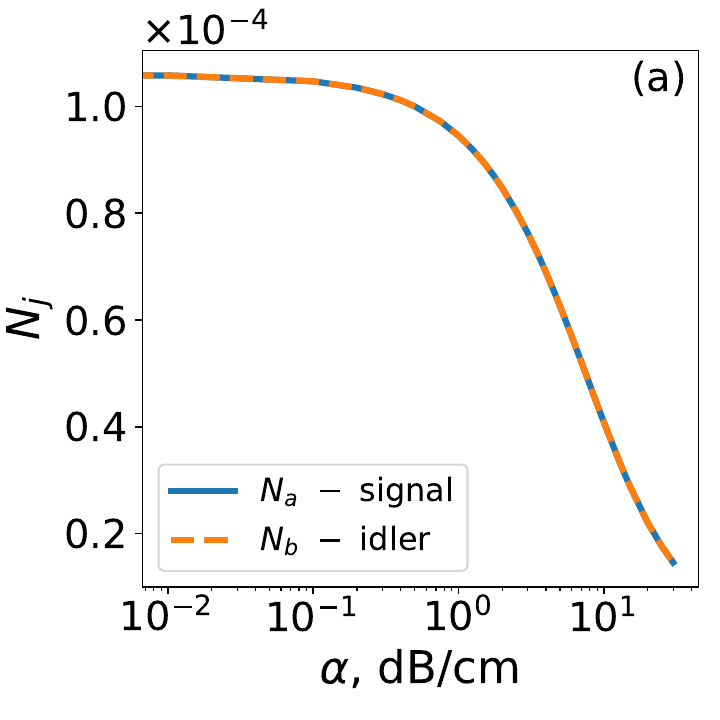 } 
    \includegraphics[width=0.245\linewidth]{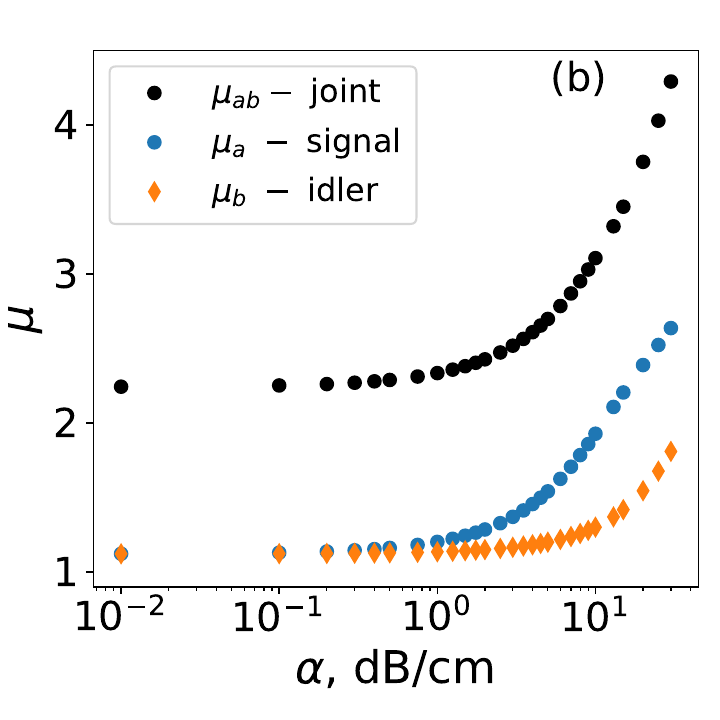 } 
    \includegraphics[width=0.245\linewidth]{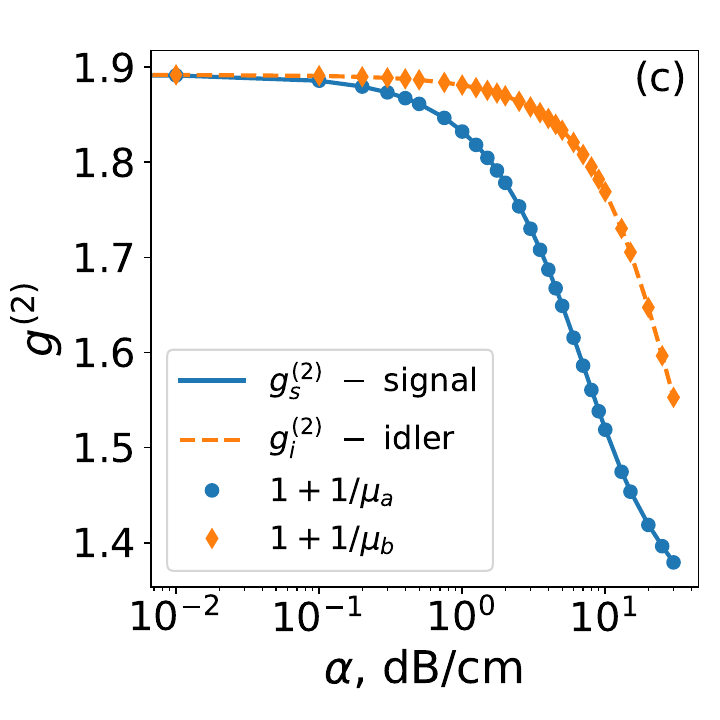 } 
    \includegraphics[width=0.245\linewidth]{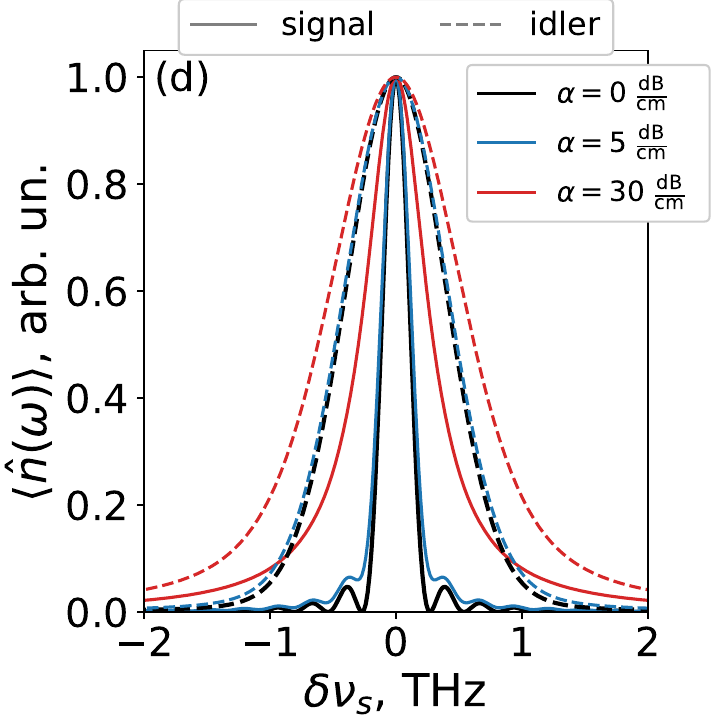 } \\
    \includegraphics[width=0.245\linewidth]{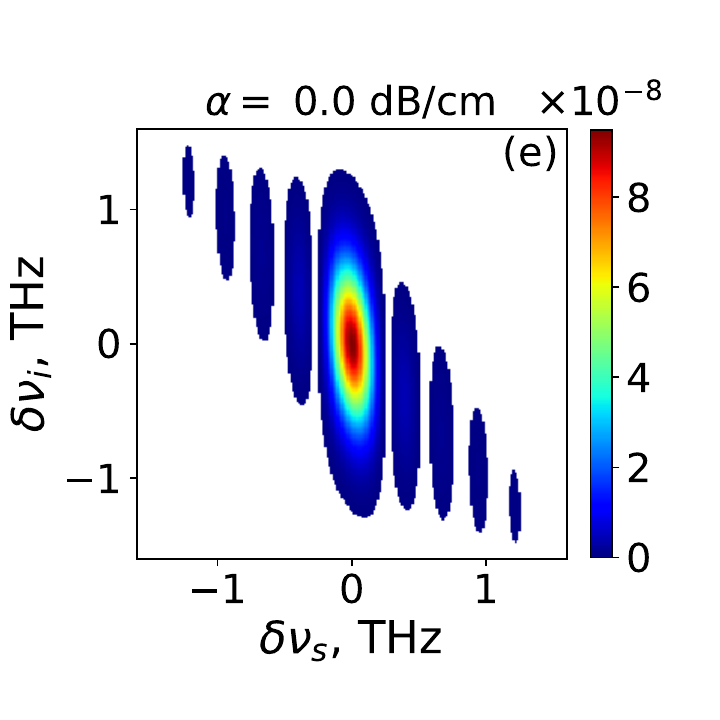 } 
    \includegraphics[width=0.245\linewidth]{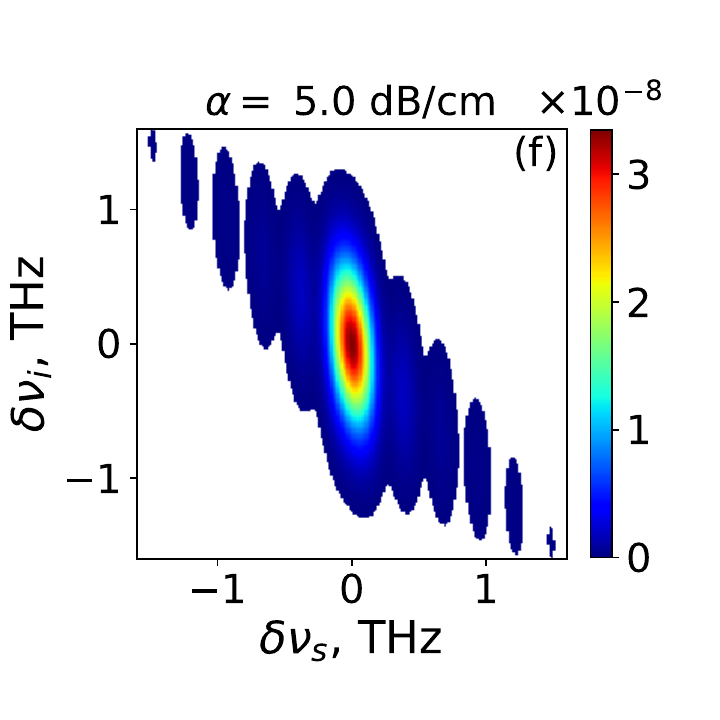 } 
    \includegraphics[width=0.245\linewidth]{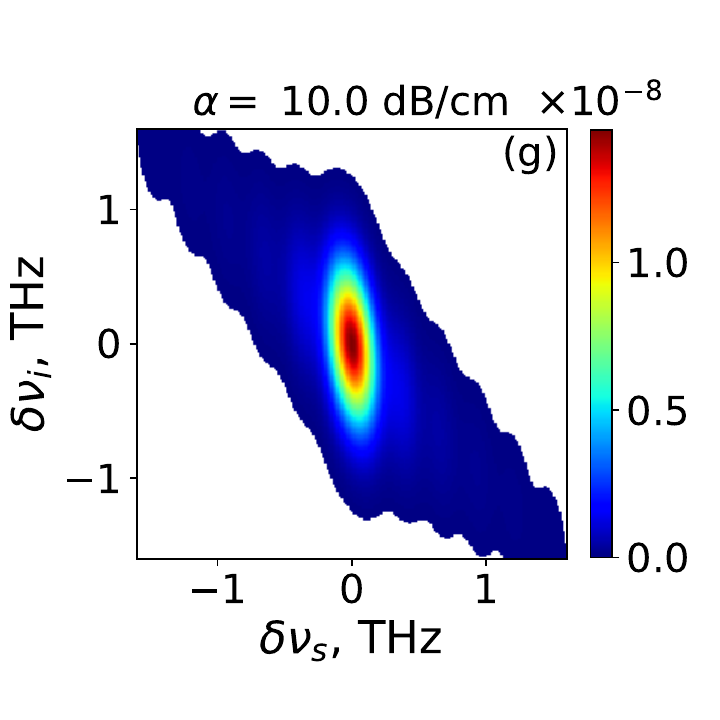 } 
    \includegraphics[width=0.245\linewidth]{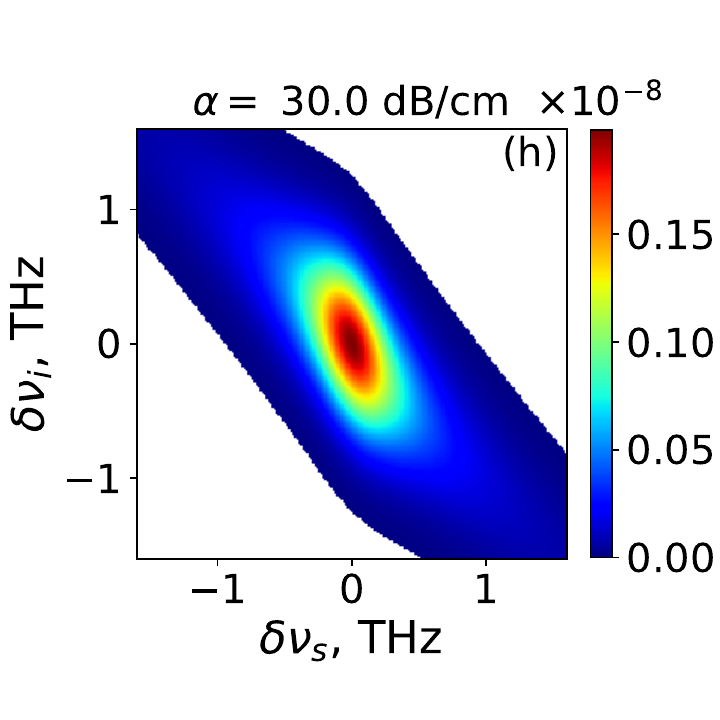 } 
    \caption{
    (a) Total number of photons for the signal and idler fields as a function of the loss coefficient $\alpha$;
    (b) number of occupied Mercer-Wolf modes $\mu_{ab}$, $\mu_a$ and $\mu_b$ for the joint system, signal and idler subsystems, respectively;
    (c) the measurement-based second-order correlation function $g^{(2)}_j$ for the signal and idler fields as a function of $\alpha$;
    (d) normalized signal and idler spectra for the lossless PDC $\alpha=0$~dB/cm and lossy PDC with $\alpha=5$~dB/cm and $\alpha=30$~dB/cm;
    (e-h) JSI for the lossless PDC with $\alpha=0$~dB/cm and lossy PDC with $\alpha=5$~dB/cm, $\alpha=10$~dB/cm and $\alpha=30$~dB/cm, respectively. 
    The white region corresponds to the values of the JSI below 0.4\% of its maximal value.
    In (d-h) $\delta \nu$ is the detuning from the central frequency of PDC $\nu_p/2$.
    } 
    \label{fig_2_general_results}
\end{figure*}
In contrast to the standard descriptions, our approach is based on the framework of Gaussian states, to which the PDC state belongs, which allows us to work beyond the two-photon approximation taking into account all Fock-state contributions.
The spatial Langevin and master equations obey causality (spatial ordering), which gives us the accurate dynamics of the generated field.
The presence of all Fock-state contributions in the solution allows us not only to obtain the proper values of $g^{(2)}$ but also to apply the derived formulas to the intermediate- and high-gain regimes.

\section{Numerical results and discussion}
\label{sec_results}

To generate frequency-degenerate type-II PDC, we consider a $1$~cm long waveguide with manually defined dispersion and losses.
As pump, we use a Gaussian pulse with a full width at half maximum of $\Delta\tau = 0.5$~ps and a central wavelength of $\lambda_p = 755$~nm.

So far as we consider long pulses with narrow spectra, we can limit ourselves to the first-order refractive index expansion for the pump, signal, and idler waves, i.e., we do not consider group-velocity dispersion or chirp in the waveguide.
In this case the refractive index for each field is taken to be
\begin{equation}
    n(\omega) = n(\omega_0) + \dfrac{\omega-\omega_0}{\omega_0}\bigg[\dfrac{c}{v_g(\omega_0)} - n(\omega_0)\bigg],
\end{equation}
where $c$ is the speed of light and $v_g$ is the group velocity.
In order to model a waveguide, we choose the following parameters: the pump refractive index  $n_p = n(\omega_p) = 1.9$ and group velocity $v_g^p = 0.9 c/n_p $, the signal 
 refractive index $n_s = n(\omega_p/2) = 1.9$ and group velocity $v_g^s = 0.95 v_g^p$, 
the idler refractive index $n_i = n(\omega_p/2) = 1.8$ and group velocity $v_g^i = v_g^p$.
Note, that here we study the regime of group-velocity-matching between the pump and idler waves.
Experimentally, such type of phase-matching was studied in, e.g., Ref.~\cite{Xin_2022}.
The quasi-phase-matching is obtained with $k_{QPM} = \frac{\omega_p}{2c} (2 n_p - n_s - n_i)$.

In the numerical computations we assume equal  losses for the signal and idler fields $\alpha_s=\alpha_i=\alpha$. 
The initial state and the state of the environment are taken to be vacuum. 
The pump is assumed to be non-scattered ($\alpha_p = 0$).
Below we study the case of spontaneous PDC with $\Gamma L \ll 1$ and $\braket{\hat{n}} \ll 1$.
 
\subsection{Spectral properties}
\label{sec_results_spectrum}

In Fig.~\ref{fig_2_general_results}, the numerical results for the considered waveguide are presented.
The number of photons for the signal and idler fields as a function of the loss parameter $\alpha$ is shown in Fig.~\ref{fig_2_general_results}(a).
For lossless PDC, the average number of PDC photons per pulse reads $N={N}_a+{N}_b=2.1\cdot 10^{-4}$, which corresponds to the spontaneous regime of PDC.
As expected, the number of photons decreases with the increasing loss coefficient.  
So far as we consider equal losses for the signal and idler fields, the dependencies for the signal and idler fields coincide. 

In contrast to the number of photons, the effective number of occupied PDC modes increases with the loss coefficient (see Fig.~\ref{fig_2_general_results}(b)).
For lossless PDC $\mu_{ab} = 2.2$, while its change for $\alpha < 0.5$~dB/cm is less than $2\%$.
Starting from $\alpha \approx 1$~dB/cm, the number of modes increases significantly.

In addition, the dependencies of the effective number of occupied modes  for the signal ($\mu_a$) and idler ($\mu_b$) subsystems are shown in Fig.~\ref{fig_2_general_results}(b) and illustrate the same tendency of increasing numbers of modes with the loss coefficient.
However, despite considering equal losses in both channels, these dependencies are different, which indicates different spectral and temporal structures of signal and idler subsystems.
The modification of the mode structure of PDC is also illustrated in Fig.~\ref{fig_2_general_results}(c), where the $g_{s,i}^{(2)}$ for the signal and idler fields are presented (Eqs.~\eqref{eq_g2func_meas} and \eqref{eq_g2func_modes} ).

In order to study the influence of losses on the spectral properties of PDC, the spectra of signal and idler fields are shown in Fig.~\ref{fig_2_general_results}(e) for different amounts of losses: lossless PDC with  $\alpha=0$~dB/cm, $\alpha=5$~dB/cm and $\alpha=30$~dB/cm.
Despite the noticeable amount of losses of $\alpha=5$~dB/cm, the spectra do not differ significantly from the lossless PDC: only the visibility of oscillations in the signal spectrum decreases, while the spectral width remains almost the same.

For large losses, the difference in the spectrum becomes more prominent: The oscillations in the signal spectrum disappear and the spectrum broadens in comparison to the lossless case.
Qualitatively, one can understand this as an effective reduction of the length of the nonlinear medium. 
This makes sense as high losses mean that photons, generated at the beginning of the waveguide are most likely to be scattered. 
Therefore, photons exiting the system are significantly more likely to have been generated at the end of the medium.
These effects are also revealed in the JSI. 
In Fig.~\ref{fig_2_general_results}(e-h), the JSI for different losses are shown.


\subsection{Temporal properties}
\label{sec_results_temporal}

\begin{figure}[h!]
    \includegraphics[width=0.99\linewidth]{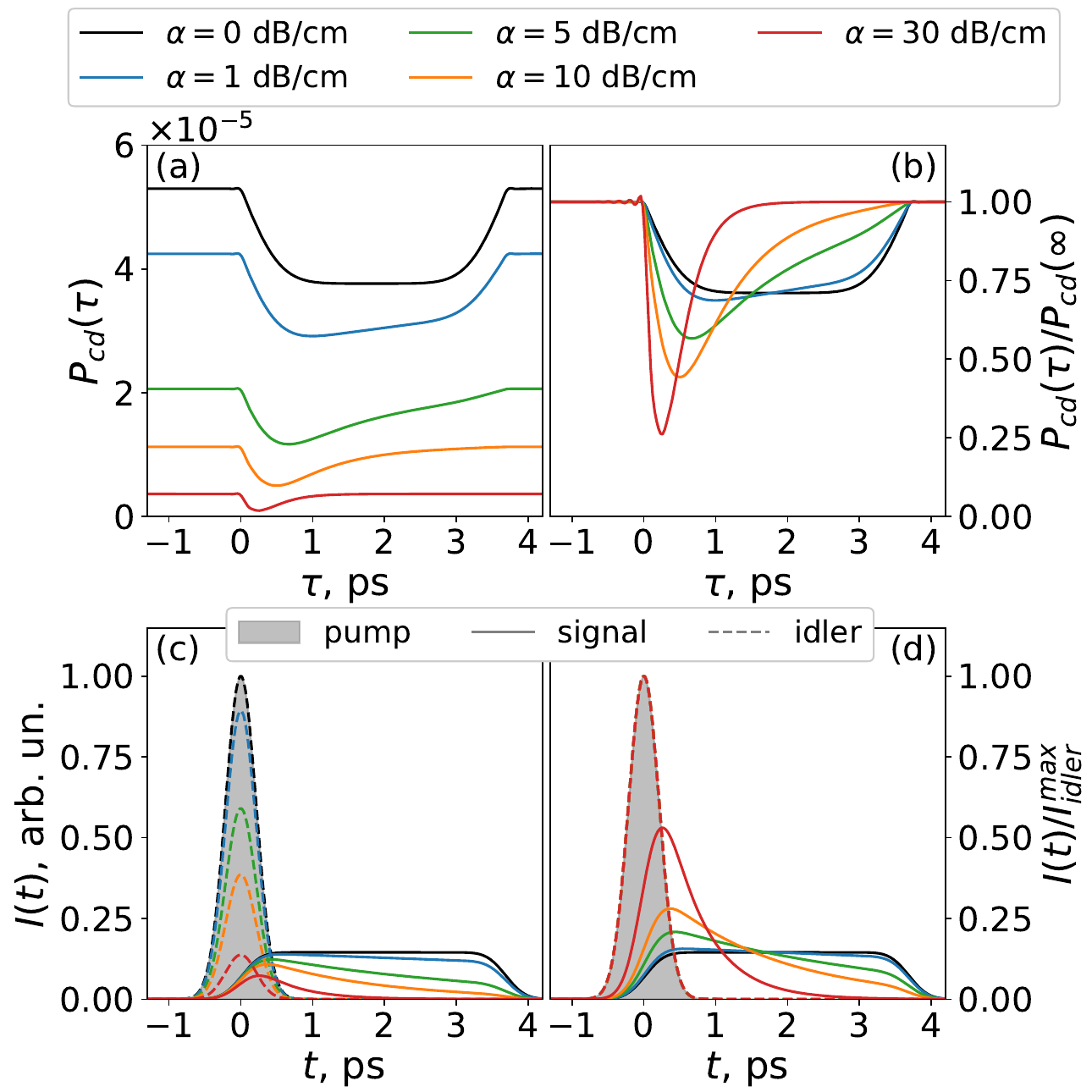 } \\
    \caption{(a, b) The absolute and normalized HOM interference patterns; 
    (c) temporal profiles of the signal and idler fields; (d)~signal and idler temporal profiles normalized to the maximal value of the idler field.
    Different colors correspond to different values of $\alpha$; different line-styles in (c,d) correspond to signal and idler field. 
    The filled area represents the temporal profile of the pump field.
    }
    \label{fig_3_hom_results}
\end{figure}

In Fig.~\ref{fig_3_hom_results}(a,b), the HOM patterns are presented. 
First, in Fig.~\ref{fig_3_hom_results}(a), the absolute values of coincidence probabilities between detectors are given for lossless PDC and lossy PDC.
The increase in $\alpha$ leads to a decrease of the maximal coincidence probability.
In addition, the shape of the interference pattern is changed, which is explicitly demonstrated for the normalized coincidence probabilities in Fig.~\ref{fig_3_hom_results}(b).
As the losses increase, one can notice that the visibility of the HOM interference increases while the temporal width of the HOM dip decreases.

To explain all observed effects, the temporal profiles of the signal and idler fields are shown in Fig.~\ref{fig_3_hom_results}(c,d).
Since the group velocity of the pump field equals the group velocity of the idler field, the temporal profiles of the pump and idler fields coincide.
In turn, the signal field is slower.
During the pump propagation along the waveguide, the generated signal photons are delayed with respect to the pump pulse, which results in a temporal profile of the signal field that has a large plateau. 
The earlier the signal photons are generated, the more delayed they are with respect to the pump.
This effect is known as a temporal walk-off~\cite{Agrawal_book}. 

In the presence of losses, the photons are scattered, reducing the intensity of the PDC fields for both the signal and the idler fields.
The idler pulse profile does not change significantly, while the signal pulse shape reveals a skew.
This skew can be interpreted in the following manner: The amount of lost photons is proportional to the traveled distance inside the scattering medium. 
The photons generated at the beginning of the waveguide are more likely to be scattered, compared to the photons generated in the middle and in the end of the waveguide.
Due to the temporal walk-off, we observe this effect as a skew in the temporal profile of the signal field. 
On the opposite, the idler temporal profile completely coincides with the pump and its shape does not change with losses. 
Nevertheless, for $\alpha_s=\alpha_i$, it reveals the same amount of losses as the signal field, which can be noticed in Fig.~\ref{fig_2_general_results}(a) showing the total number of photons in each subsystem.

Despite such a destructive behavior of losses, they increase the overlap between the resulting signal and idler fields (see Fig.~\ref{fig_3_hom_results}(d)).
The increased similarity between the temporal profiles of the signal and idler photons leads to increasing visibility of the HOM dip.
Higher visibility is usually interpreted as better biphoton indistinguishability, so high internal losses can reduce the difference of temporal profiles of signal and idler fields and make them more indistingiushable.

\subsection{Determination of losses}
\label{sec_results_new_method}

\begin{figure}
    \includegraphics[width=.49\linewidth]{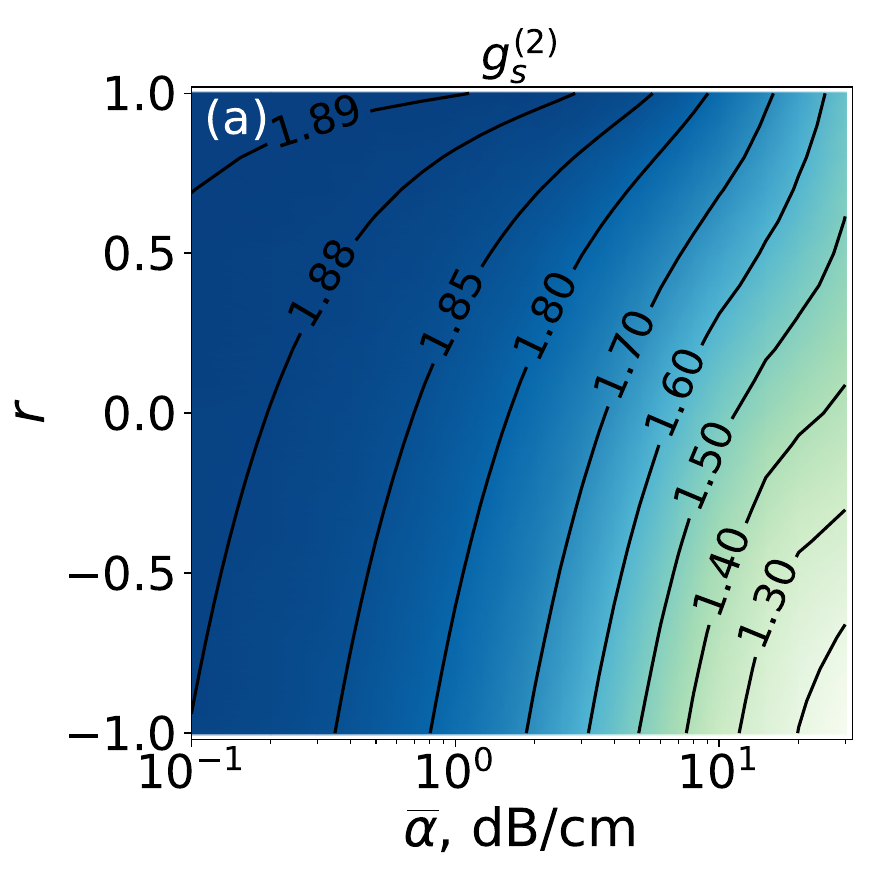} 
    \includegraphics[width=.49\linewidth]{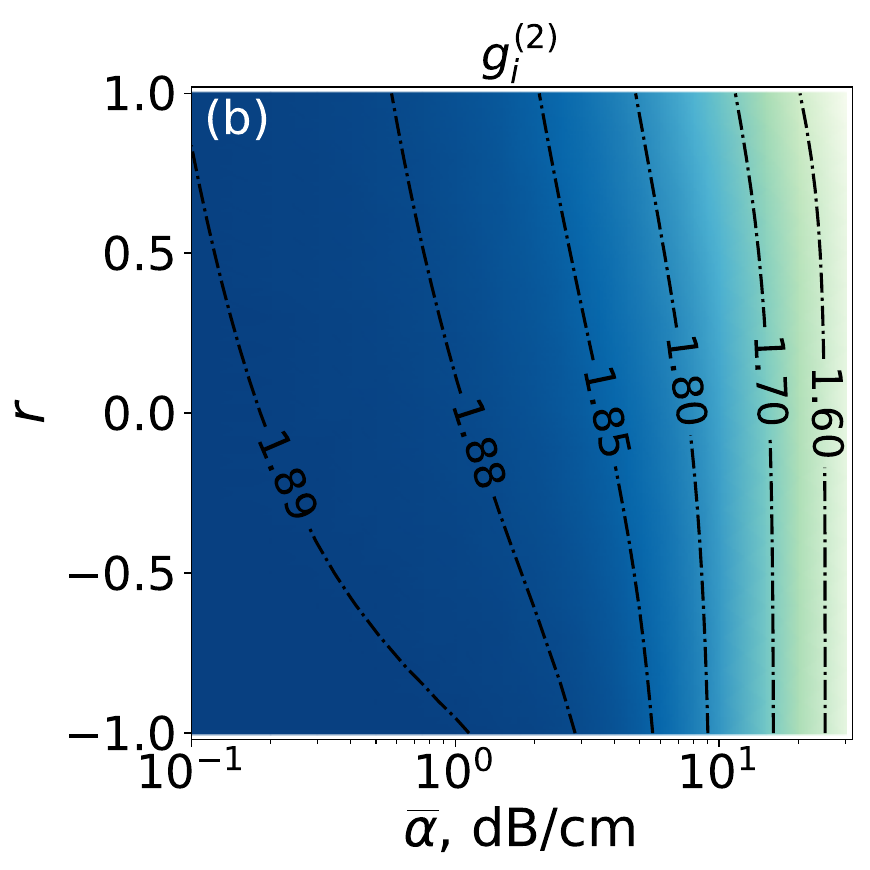 } \\
    \includegraphics[width=.49\linewidth]{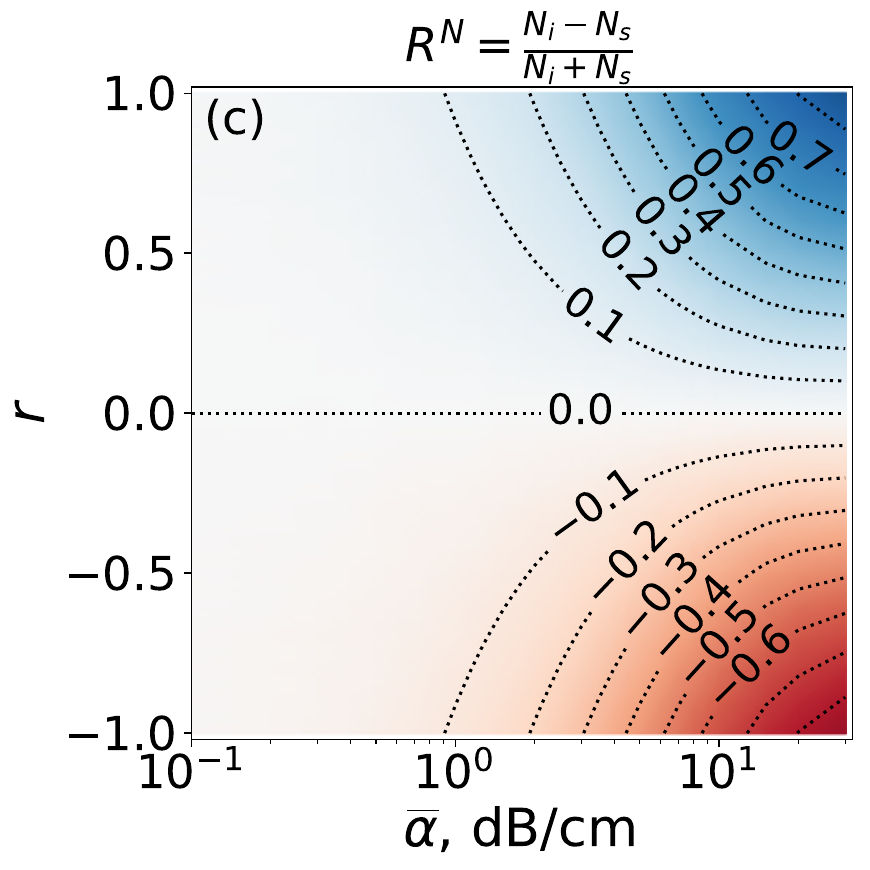} 
    \includegraphics[width=.49\linewidth]{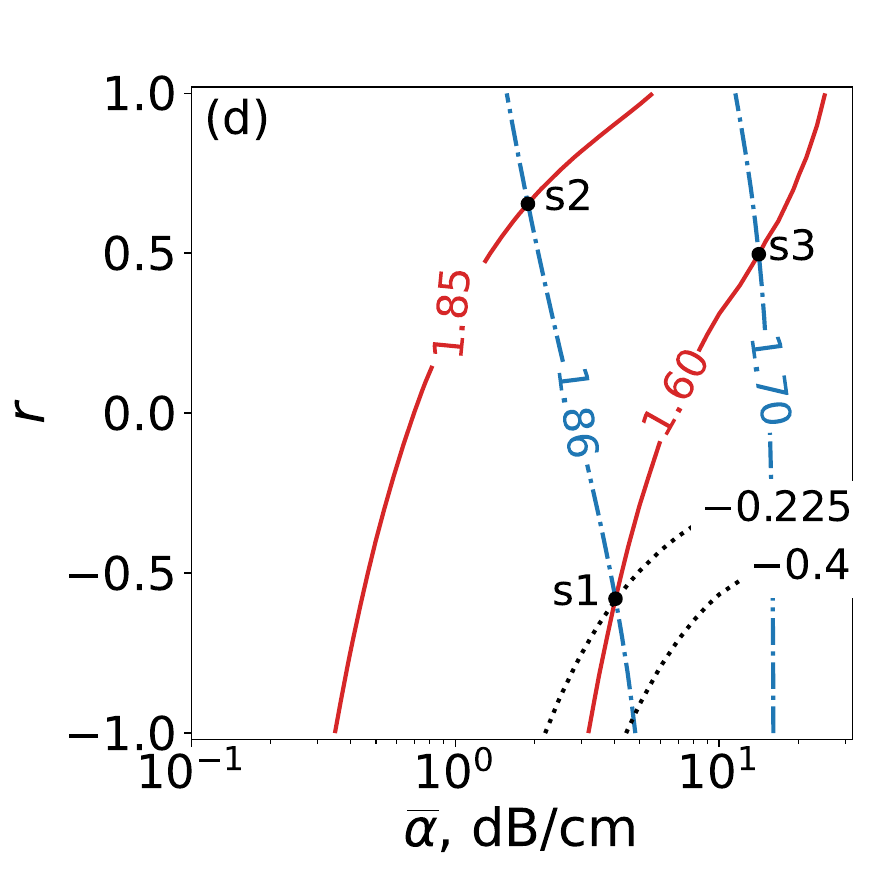 } 
    \caption{ 
        (a, b, c) The dependencies of $g^{(2)}_s$, $g^{(2)}_i$ and $R^N$ on $\bar\alpha$ and $r$ , respectively. The waveguide dispersion and the pump profile are the same as in Section~\ref{sec_results}. 
        (d) 
        The black dots indicate the intersection of two isolines, $g^{(2)}_s$ (red, solid) and $g^{(2)}_i$ (blue, dash-dotted), that correspond to the `measured' $g^{(2)}$ values. 
        The black dotted curves depict the isolines of $R^N$.        
        \\
            (point s1) $g^{(2)}_s = 1.6$ and $g^{(2)}_i = 1.86$, which gives the estimated values $\bar\alpha_1 = 4.0 $~dB/cm and $r_1 = -0.57$, \\
            (point s2) $g^{(2)}_s = 1.85$ and $g^{(2)}_i = 1.86$, which gives the estimated values $\bar\alpha_2 = 1.9$~dB/cm and $r_2 = 0.65$, \\
            (point s3) $g^{(2)}_s = 1.6$ and $g^{(2)}_i = 1.7$, which gives the estimated values $\bar\alpha_3 = 14$~dB/cm and $r_3 = 0.49$.
        }
    \label{fig_4_method}
\end{figure}

The results of the previous subsection demonstrate that the values of $g^{(2)}$ for the signal and idler modes depend differently on the internal losses $\alpha$, even if the losses are the same for the signal and idler channels.
Since external frequency-independent losses (transmission losses) do not change the value of normalized second-order correlation function (see Appendix~\ref{appendix_g2}),  
the difference in $g^{(2)}$ between the signal and idler fields can be used as an
indicator of the internal waveguide losses. 
This section presents how measured values of $g^{(2)}$ can be applied to experimentally determine the internal losses of a waveguide. 
 
Let us assume a waveguide with the known dispersion and a general case of different frequency-independent losses  $\alpha_s \neq \alpha_i $, which are unknown and should be determined in the experiment.
The losses can be parametrized as  
\begin{equation}
    \bar\alpha = \dfrac{\alpha_s+\alpha_i}{2}, ~ ~ ~ ~ ~   r = \dfrac{\alpha_s-\alpha_i}{\alpha_s+\alpha_i}. 
\end{equation}

So far as we know the dispersion of the waveguide, the theoretical values of $g^{(2)}_s(\bar\alpha, r)$, $g^{(2)}_i(\bar\alpha, r)$, and the relative number of photons at the waveguide output 
\begin{equation}
    R^N(\bar\alpha, r) = \dfrac{N_i-N_s}{N_s+N_i}
\end{equation}
can be calculated as functions of $\bar\alpha$ and $r$.
If the behavior of the fixed isolines $g^{(2)}_s$, $g^{(2)}_i$, and $R^N$ are different in the parameter space (see Fig.~\ref{fig_4_method}(a,b,c)), their intersection allows us to estimate the values of $\bar\alpha$ and~$r$. 
Summing up, for a known waveguide, the internal losses can be experimentally determined from the measured values of $g^{(2)}_s$, $g^{(2)}_i$ and $R^N$.

In general, the intersection can be determined only from the values of $g_s^{(2)}$ and $g_i^{(2)}$, while the value of $R^N$ is optional.
Indeed, the second-order correlation function is insensitive to the external losses (transmission and detection) and the measurements of $g_s^{(2)}$ and $g_i^{(2)}$ can be performed with high accuracy. 
In contrast, the knowledge of external losses is important for the correct measurement of a relative number of photons $R^N$, which can sometimes be challenging.
However, if the external losses are correctly estimated, the measured value of $R^N$ can verify that the theoretical model is consistent with measurements.

In Fig.~\ref{fig_4_method}(a,b,c), the dependencies of $g^{(2)}_s$, $g^{(2)}_i$, and $R^N$ are presented for the waveguide dispersion and the pump profile defined at the beginning of the Section~\ref{sec_results}.
In Fig.~\ref{fig_4_method}(d) we present three examples `s1', `s2', and `s3' that correspond to three `measurements' of $g^{(2)}_s$ and $g^{(2)}_i$. 
For each fixed `measured' value of $g^{(2)}_s$ and $g^{(2)}_i$, there are two theoretically calculated isolines, (red, solid) and (blue, dash-dotted), respectively. 
Such isolines have an intersection point (black circles) that defines the amount of internal losses.   

In Fig.~\ref{fig_4_method}(d) we also present two isolines for a relative number of photons $R^N$  (black, dotted). 
For our waveguide,  if the external losses are correctly accounted and the `measured' values of correlation functions read $g^{(2)}_s = 1.6$ and $g^{(2)}_i = 1.86$, one should experimentally obtain  $R^N=-0.225$.
If for the given values of $g^{(2)}_s$ and $g^{(2)}_i$ and correctly accounted external losses we measure another value of $R^N$ which is outside the measurement error (no joint intersection point for three isolines), for example, $R^N=-0.4$, then this means that our 
prior knowledge about the waveguide is not correct.

The dependencies of $ g^{(2)}_s $ and $g^{(2)}_i$ on losses are defined by the waveguide dispersion and should be studied for each particular case individually.
Indeed, the presence of higher spatial modes, frequency-dependent losses, or waveguide imperfections can significantly change the properties of the generated SPDC field and $ g^{(2)} $ behavior.
In addition, note that the equal values of $ g^{(2)}_s = g^{(2)}_i$ do not guarantee the absence of internal losses. 
Therefore, for each specific practical use, the proposed method should be additionally elaborated and include all these factors, what is outside of the scope of this paper.

\section{Conclusions}
\label{sec_conclusion}

In this work, we examine theoretically the spectral and temporal properties of low-gain broadband PDC generated in a lossy waveguide.
Our theoretical approach is based on the formalism of Gaussian states and the Langevin equation and is adjusted for the weak parametric down-conversion process and photon-number unresolved detection.

Using the example of frequency-degenerate type-II PDC generated under the pump-idler group-velocity-matching condition, we show how internal losses of nonlinear waveguides change the properties of the generated light. 
We demonstrate that the influence of internal losses on the spectral profiles of the generated field is weak. 
However, the Hong-Ou-Mandel interference strongly depends on losses: the Hong-Ou-Mandel  dip may increase with losses, while the correlation time decreases, which is explained in terms of the temporal profiles of the signal and idler fields.

One of the most important results is the dissimilar dependence of the number of modes of the signal and idler field on the internal losses (even when losses are equal). 
Such behavior becomes apparent in the second-order correlation functions of the signal and idler fields, which can be easily detected in experiments.
Based on this effect, we propose a new method for the experimental determination of internal losses in nonlinear waveguides. 
The presented method is based on a prior knowledge of the waveguide mode structure, which can be obtained either by theoretical calculations based on the waveguide geometry.
We believe that the results obtained in our work can be directly applied to experiments and will strongly improve the characterization of nonlinear waveguides.

\begin{acknowledgments}
This work is supported by the `Photonic Quantum Computing' (PhoQC) project, which is funded by the Ministry for Culture and Science of the State of North-Rhine Westphalia.
\end{acknowledgments}

\appendix

\section{Covariance matrix}
\label{appendix_covariance_matrix}

A covariance matrix is a real positive-definite symmetric matrix of the second-order moments of the quadrature operators~\cite{Weedbrook_2012}.
In this paper, we define quadrature operators ($\hbar=2$) as $\hat{q}_i = \hat{c}^\dagger_i + \hat{c}_i $ and $\hat{p}_i =  i(\hat{c}^\dagger_i - \hat{c}_i)$ with the and commutation relations $[\hat{q}_n, \hat{p}_m] = 2i\delta_{nm}$.

For non-displaced quantum states with $\braket{\hat{c}_i} = 0$ and $\braket{\hat{x}_i } = 0$, the elements of the covariance matrix $\sigma$ are given by
\begin{equation}
  \sigma_{ij} = \dfrac{\braket{\hat{x}_i\hat{x}_j + \hat{x}_j\hat{x}_i }}{2} ,
  \label{eq_covariance}
\end{equation}
where $\hat{x}_i$ are the elements of the vector $\hat{\mathbf{x}}  = (\hat{q}_1, \hat{q}_2, \dots, \hat{q}_{2N}, \hat{p}_1, \hat{p}_2, \dots, \hat{p}_{2N})^T$.
Having the second-order correlators $\braket{\hat{c}^\dagger_i\hat{c}_j}$ and $\braket{\hat{c}_i\hat{c}_j}$, the elements of the matrix $\sigma$ are given by
\begin{align}
    \label{eq_covariance_expressions_1}
    \braket{ \hat{q}_i \hat{q}_j} = 
    \delta_{ij} 
     + 2 \Big(\text{Re}[\braket{\hat{c}^\dagger_i\hat{c}_j}]
     + \text{Re}[\braket{\hat{c}_i\hat{c}_j}] \Big), \\
    \braket{ \hat{p}_i \hat{p}_j}  = 
    \delta_{ij} 
     + 2 \Big(\text{Re}[ \braket{\hat{c}^\dagger_i\hat{c}_j}]
     - \text{Re}[\braket{\hat{c}_i\hat{c}_j}] \Big), \\
    \dfrac{ \braket{ \hat{p}_i \hat{q}_j +  \hat{q}_j \hat{p}_i} }{2} =  2 \Big(\text{Im}[\braket{\hat{c}_i\hat{c}_j}] -\text{Im}[ \braket{\hat{c}^\dagger_i\hat{c}_j} ]   
       \Big).
    \label{eq_covariance_expressions_3}
\end{align}

To build the covariance matrix of the joint signal-idler system $\sigma^{ab}$, the $ \braket{\mathbf{\hat{c}^\dagger \hat{c} }} = \mathcal{D}$ and $ \braket{\mathbf{\hat{c} \hat{c} }} =\mathcal{C}$ from the main text.
In turn, a covariance matrix $\sigma^{a}$ for the signal subsystem is defined by the matrices $ \braket{\mathbf{\hat{c}^\dagger \hat{c} }} = \braket{\mathbf{\hat{a}^\dagger \hat{a}} } $ and $\braket{\mathbf{\hat{c}\hat{c} }} =  \braket{\mathbf{\hat{a} \hat{a}} } $.
 

\section{Second-order correlation function and external losses}
\label{appendix_g2}

The normalized second-order correlation function is insensitive to external frequency-independent losses.
Indeed, for the field $\hat{E}^{(+)}(t)$, the losses can be introduced via a virtual beamsplitter with transmission coefficient~$T$.
If the losses are the same for all frequencies, the field transformation has the form $\hat{E}^{(+)}(t) \rightarrow \sqrt{T} \hat{E}^{(+)}(t)$.
By substitution the transformed field into Eq.~\eqref{eq_g2func_def}, the factors $\sqrt{T}$ cancel out, what keeps the function $g^{(2)}(t_1, t_2, t_3, t_4)$ unchanged.

As result, the values $g^{(2)}_s$ and $g^{(2)}_i$ remain unchanged in the presence of frequency-independent external losses (transmission and detection losses).
As long as for lossless PDC $g^{(2)}_s = g^{(2)}_i$, a difference $g^{(2)}_s-g^{(2)}_i$ can indicate the presence of internal losses during the PDC process.

\bibliography{lit_losses_nondegen.bib}

\end{document}